\documentclass[a4paper]{spie}  

\usepackage{amsmath,amsfonts,amssymb}
\usepackage{graphicx}
\usepackage[colorlinks=true, allcolors=blue]{hyperref}

\usepackage[perpage]{footmisc}

\title{The Virtual Observatory Ecosystem Facing the European Open Science Cloud}

\author[a]{Marco Molinaro}
\author[b]{Mark Allen}
\author[b]{Fran\c{c}oise Genova}
\author[b]{Andr\'{e} Schaaff}
\author[c]{Margarida Castro Neves}
\author[c]{Markus Demleitner}
\author[a]{Sara Bertocco}
\author[d]{Dave Morris}
\author[b]{Fran\c{c}ois Bonnarel}
\author[d]{Stelios Voutsinas}
\author[e]{Catherine Boisson}
\author[a]{Giuliano Taffoni}

\affil[a]{INAF - Osservatorio Astronomico di Trieste, via G.B. Tiepolo 11, 34143 Trieste, Italy}
\affil[b]{CNRS - Centre de Donn\'{e}es astronomiques de Strasbourg, Strasbourg, France}
\affil[c]{Universit\"{a}t Heidelberg, Astronomisches Rechen Institut, Heidelberg, Germany}
\affil[d]{University od Edinburgh, Edinburgh, Scotland, UK}
\affil[e]{Observatoire de Paris - Meudon, Paris, France}

\authorinfo{Further information: marco.molinaro@inaf.it}

\begin{document} 
\maketitle

\begin{abstract}
The International Virtual Observatory Alliance (IVOA) has developed and built, in the last
two decades, an ecosystem of distributed resources, interoperable and based upon open 
shared technological standards. In doing so the IVOA has anticipated, putting into practice
for the astrophysical domain, the ideas of FAIR-ness of data and service resources and the
Open-ness of sharing scientific results, leveraging on the underlying open standards
required to fill the above. In Europe, efforts in supporting and developing the ecosystem
proposed by the IVOA specifications has been provided by a continuous set of EU funded
projects up to current H2020 ESCAPE ESFRI cluster. In the meantime, in the last years,
Europe has realised the importance of promoting the Open Science approach for the research
communities and started the European Open Science Cloud (EOSC) project to create a
distributed environment for research data, services and communities. In this framework the
European VO community, had to face the move from the interoperability scenario in the 
astrophysics domain into a larger audience perspective that includes a cross-domain FAIR
approach. Within the ESCAPE project the CEVO Work Package (Connecting ESFRI to EOSC through
the VO) has one task to deal with this integration challenge: a challenge where an 
existing, mature, distributed e-infrastructure has to be matched to a forming, more general
architecture. CEVO started its works in the first months of 2019 and has already worked on
the integration of the VO Registry into the EOSC e-infrastructure. This contribution 
reports on the first year and a half of integration activities, that involve applications, 
services and resources being aware of the VO scenario and compatible with the EOSC architecture.
\end{abstract}


\section{INTRODUCTION}
\label{sec:intro}  

ESCAPE Work Package 4 \textit{Connecting ESFRI projects to EOSC through VO framework} (CEVO, see 
also Sec.~\ref{subsec:cevo}), 
plans to make the seamless connection of ESFRI and other astronomy and astroparticle research 
infrastructures to the EOSC through the Virtual Observatory (VO) framework. CEVO Task 4.1 (see 
Sec.~\ref{subsec:cevot41}) has 
been investigating and applying solutions to integrate data and service resources, based 
upon the IVOA architecture (see Sec.~\ref{sec:ivoa}), into the EOSC (see Sec.~\ref{sec:eosc}) 
one. 

Progress so far has worked out more easily on the technical 
side, where the technologies involved have proved to be compatible, like it is the case of 
integrating the IVOA Resource Registry (Sec.~\ref{subsubsec:registry}). The developments on 
other aspects have been slower due to 
the challenges of interfacing with the evolving structure of the EOSC platform, for instance for 
service provision (Sec.~\ref{subsubsec:onboarding}), or due to not-yet-provided features of the 
architecture, for instance for the 
integration of semantics and vocabularies (Sec.~\ref{subsubsec:semantics}).
Participation in relevant EOSC project 
events has been of great help in identifying interfaces and 
contacts to allow the integration activities.

Connecting to other EOSC-related projects (activity done in connection with CEVO Task 4.2) allowed 
describing the Open and FAIR\cite{sdata.2016.18} approach, envisioned in the IVOA, to other research domains and to 
projects whose purpose is to define building blocks for FAIR data sharing. 
This helped with the clarification of concepts with 
respect to data policies, licensing, user identification and permission granting. These concepts
are a major consideration when providing resources but should not impede the easy access to public 
data holdings that form the predominant case in astrophysics.

Task activities involved connections to the other ESCAPE Work Packages (see 
Sec.~\ref{subsec:crossWPintercation}):
\begin{itemize}
    \item to make available a portfolio of software and services through EOSC, also through containerisation (WP3 - OSSR);
    \item to contribute to the EOSC Hybrid Cloud with technical standard solutions for the ESCAPE Science Platform (WP5 - ESAP);
    \item also considering the need for this latter service to have a standardised access to the Data Lake solution (WP2 - DIOS).
\end{itemize}
The outcome of the investigations done during the first half of the project (that updates the 
initial report ADASS XXIX\cite{2019arXiv191108205M}) is analysed in 
Sec.~\ref{sec:analysis} and the open questions and challenges in Sec.~\ref{sec:next}. 

\section{ESCAPE}

\subsection{Project Overview}
\label{sec:escape}

ESCAPE (\textit{European Science Cluster of Astronomy \& Particle physics ESFRI research 
infrastructures})\footnote{\url{https://projectescape.eu/}} addresses the Open Science challenges shared by ESFRI facilities (SKA, CTA, 
KM3Net, EST, ELT, HL-LHC, FAIR) as well as other pan-European research infrastructures (CERN, ESO, 
JIVE) in astronomy and particle physics. 

The investigated solutions shall:
\begin{itemize}
    \item connect ESFRI projects to EOSC ensuring integration of data and tools;
    \item foster common approaches to implement open-data stewardship; 
    \item establish interoperability within EOSC as an integrated multi-messenger facility.
\end{itemize}
To accomplish these objectives ESCAPE unites astrophysics and particle 
physics communities and their expertise in computing and data management in support of the FAIR 
principles. 
ESCAPE supports existing infrastructures such as the astronomy VO (represented by the EURO-VO 
coordination effort, Sec.~\ref{sec:eurovo}) to
connect with the EOSC. With the commitment from various ESFRI projects in the cluster, ESCAPE will 
develop and integrate the EOSC catalogue with a dedicated catalogue of open source analysis 
software and services developed by the astronomy and particle physics communities.
Through this catalogue ESCAPE will strive to provide researchers with consistent access to an integrated Open Science platform for data-analysis workflows.

\subsection{CEVO Work Package Description}
\label{subsec:cevo}

Within the Work Packages (WPs) of the ESCAPE project, WP4 \textit{CEVO} -- 
\textit{Connecting ESFRI projects to EOSC through VO framework} -- is meant to:
\begin{itemize}
    \item assess and implement the connection of the ESFRI and other astronomy Research Infrastructures to the EOSC through the Virtual Observatory framework;
    \item refine and further pursue implementation of FAIR principles for astronomy data via the use and development of common standards for interoperability including the extension of the VO to new communities, and
    \item establish data stewardship practices for adding value to the scientific content of ESFRI data archives.
\end{itemize}

ESFRI facilities from astronomy and astroparticle physics in the 
VO were already part of the collaboration between ESFRI pathfinders and European VO teams 
in the ASTERICS DADI (\textit{Data Access, Discovery and Interoperability}\footnote{\url{https://www.asterics2020.eu/work-package/dadi}}). 

\begin{figure} [ht]
   \begin{center}
   \begin{tabular}{c} 
   \includegraphics[height=5cm]{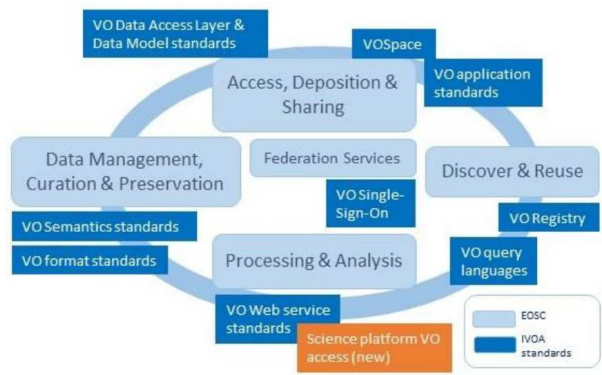}
   \end{tabular}
   \end{center}
   \caption[eosc-vo] 
   { \label{fig:eosc-vo} 
   VO mapping onto the EOSC data lifecycle.
}
   \end{figure}

The CEVO objectives build
on that background expertise to make the seamless connection of ESFRI and other astronomy research
infrastructures to the EOSC through the VO, extending them by the inclusion of new partners in 
the wider context of ESCAPE, and the new opportunities provided by EOSC. 
CEVO aims to map the VO framework to the EOSC so that the VO enabled archive services from ESFRI
will be interoperable. The diagram in Fig.~\ref{fig:eosc-vo}, inspired by the EOSC-hub
\footnote{\url{https://www.eosc-hub.eu/}} data lifecycle, shows elements 
of the
mapping of IVOA standards to different EOSC concepts in the lifecycle.
The connection of the VO registry will be key for discovery and reuse, and other standard map to 
the access, deposition and sharing of data, as
well as for data management curation and preservation.
CEVO will also enable the implementation of FAIR principles for ESFRI data via the adoption and
development of common standards for interoperability (with Task 4.2).
Another CEVO Task (Task 4.3) will pursue the application of machine learning to add value to ESFRI 
archives and will identify best practices in data stewardship.

\subsection{CEVO Task on Integrating the VO Architecture in EOSc}
\label{subsec:cevot41}

CEVO is focused on connecting ESFRI research infrastructures to the EOSC through the VO framework. 
Task 4.1 -- \textit{Integration of astronomy VO data and services into the EOSC} -- 
is dedicated to identifying the relevant domain 
specific data access, discovery and manipulation standards and integrating them with the EOSC 
platform.
To map the VO framework to the EOSC, attempting to include the VO enabled
archive services from ESFRIs in the latter, task activities will focus on four main points:
\begin{itemize}
    \item interfacing the VO framework with the EOSC;
    \item build an Astronomy portfolio of VO services;
    \item contributing to the EOSC Hybrid Cloud;
    \item containerising domain-specific services.
\end{itemize}
These integration areas have been defined into activities, reported in the CEVO Project Plan (ESCAPE 
Deliverable document D4.1\cite{escape-d41}), and their progress is also summarised in an intermediate 
analysis report (ESCAPE Deliverable document D4.4\cite{escape-d44}).
This proceeding reports them to a larger community.

\section{IVOA}
\label{sec:ivoa}

The International Virtual Observatory Alliance\footnote{\url{http://www.ivoa.net}} is an organisation, 
formed in June 2002, that focuses on the development of technological standards with the vision that
astronomical datasets and other resources should work as a seamless whole. Its members are projects that 
embrace national level communities or super-national institutes or organizations.
It currently has 21 members, plus a couple that expressed interest in joining.

\begin{figure} [ht]
   \begin{center}
   \begin{tabular}{c} 
   \includegraphics[height=8cm]{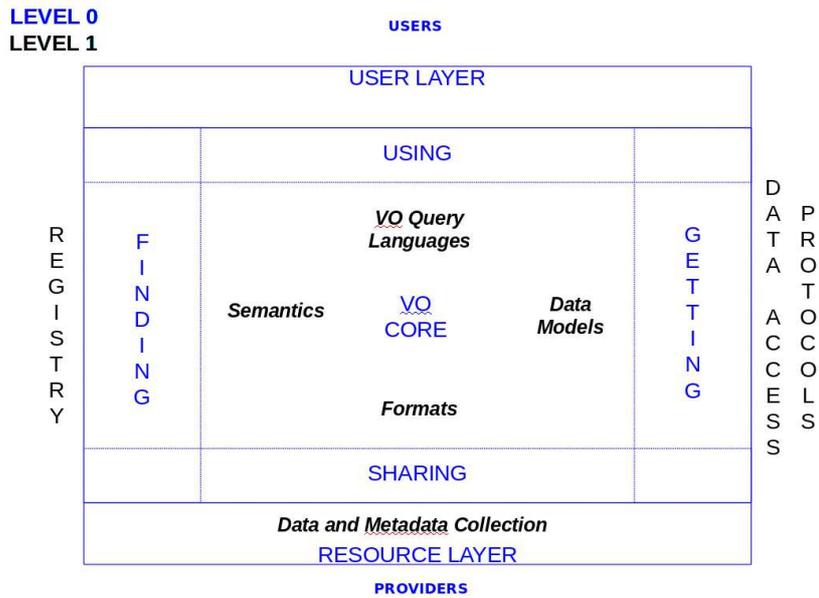}
   \end{tabular}
   \end{center}
   \caption[vo-arch] 
   { \label{fig:vo-arch} 
   Levels \textit{0} (black) and \textit{1} (blue) of the IVOA Architecture superimposed. Users, on top, find the resources they need querying the Registry, once they have the description(s) of the relevant data and service they need for their analysis, they access them by means of interfaces defined by the access protocols. All of this is standardised by IVOA Recommendations that has a core of formats, languages, models and semantic annotations to make to process interoperable.
}
\end{figure}

The architecture\cite{2010ivoa.rept.1123A} of the IVOA (briefly reported in Fig.~\ref{fig:vo-arch}) describes how the endorsed technical 
standards place themselves in the scenario of users searching for astrophysical resources, accessing them
through interoperable, standardised interfaces while being helped by formats, languages, semantic 
description and models that are shared by the VO community.

\begin{figure} [ht]
   \begin{center}
   \begin{tabular}{c} 
   \includegraphics[height=8cm]{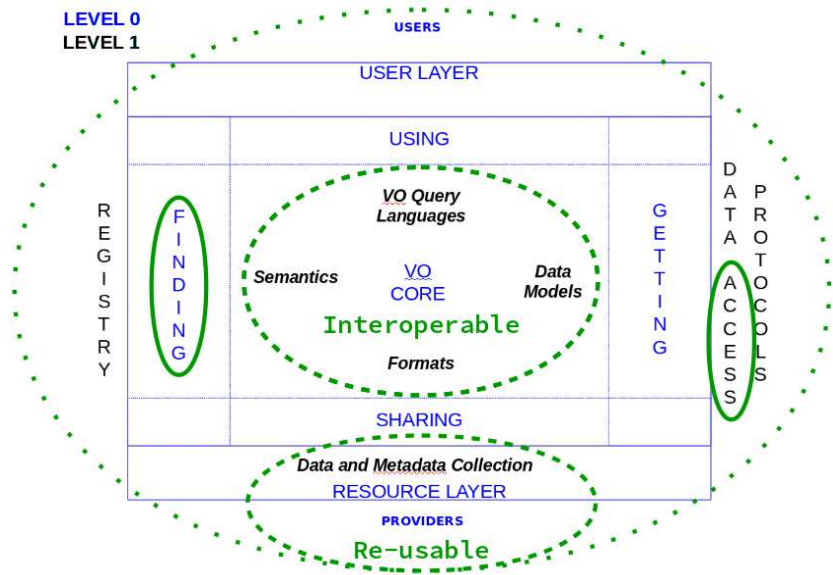}
   \end{tabular}
   \end{center}
   \caption[vo-fair] 
   { \label{fig:vo-fair} 
   Areas of the IVOA architecture highlighted with respect to their support of the FAIR principles main concepts.
}
\end{figure}

While the works of the IVOA started in 2002, the architecture itself shows quite clearly that the VO 
vision for astronomical resources embeds already the nowadays common concepts described in the FAIR 
principles (see Fig.~\ref{fig:vo-fair}).

\section{EURO-VO}
\label{sec:eurovo}

In Europe, the VO community got together at the beginning of the history of the VO and coordinated around
a sequence of \textit{EURO-VO}\footnote{\url{http://www.euro-vo.org/}} projects up to the currently 
running ESCAPE project (partly presented in 
this contribution, Sec.~\ref{sec:escape}). From the initial projects that touched directly one or another
aspect of the interoperable scenario that was the goal of the overall activities, the VO effort in Europe
became later a part of a larger cluster project that started to bring the VO expertise to other sibling 
communities: that was the case of the ASTERICS project and its Work Package DADI. Currently, the 
EURO-VO coordination is present with its members in ESCAPE and is leading the efforts in integrating the 
VO data and services within the EOSC.

\section{EOSC}
\label{sec:eosc}

The European Open Science Cloud 
(EOSC\footnote{
\url{https://ec.europa.eu/info/research-and-innovation/strategy/goals-research-and-innovation-policy/open-science/eosc_en}
}) 
is an environment for hosting and processing research data to support EU science.

The process to create the EOSC was initiated by the Commission in 2015. It aimed to develop a trusted, 
virtual, federated environment that cuts across borders and scientific disciplines to store, share, 
process and re-use research digital objects (like publications, data, and software) following FAIR 
principles.

The EOSC brings together institutional, national and European stakeholders, initiatives and data 
infrastructures to develop an inclusive open science ecosystem in Europe.

This can lead new insights and innovations, higher research productivity and improved reproducibility in 
science.

\section{PROGRESS OF INTEGRATION ACTIVITIES}
\label{sec:integration}

The activities described in this section have taken advantage of the networking activities, 
reporting and knowledge acquired by participating in relevant EOSC project events. 
Progress report is subdivided following the main activity blocks identified by the CEVO 
work plan; however, activities can be overlapping or cross the borders of the WPs 
themselves.

\subsection{VO to EOSC Interfacing}
\label{subsec:eoscinterface}

This section is further subdivided in the several aspects of the interfacing of the VO with EOSC: the registry of 
resources, resource on-boarding, and semantics vocabularies.

\subsubsection{Registry of resources}
\label{subsubsec:registry}

\begin{figure} [ht]
   \begin{center}
   \begin{tabular}{c} 
   \includegraphics[height=4cm]{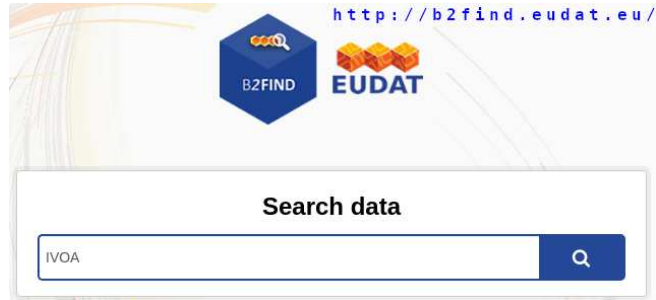}
   \end{tabular}
   \end{center}
   \caption[b2find-search] 
   { \label{fig:b2find-search} 
   EUDAT B2FIND search box, initialised with a search using the term \textit{IVOA}.
}
\end{figure}

\begin{figure} [ht]
   \begin{center}
   \begin{tabular}{c} 
   \includegraphics[height=7cm]{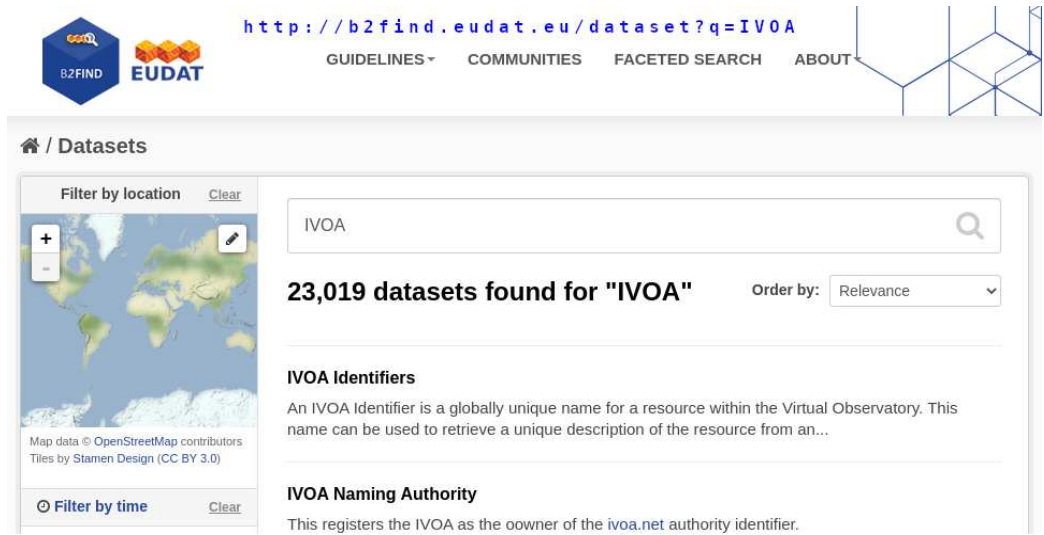}
   \end{tabular}
   \end{center}
   \caption[b2find-results] 
   { \label{fig:b2find-results} 
   B2FIND \textit{IVOA} results. The dataset count displays the actual content of the 
   IVOA Registry at the moment the inclusion took place and the snapshot was taken.
}
\end{figure}

\begin{figure} [ht]
   \begin{center}
   \begin{tabular}{c} 
   \includegraphics[height=8cm]{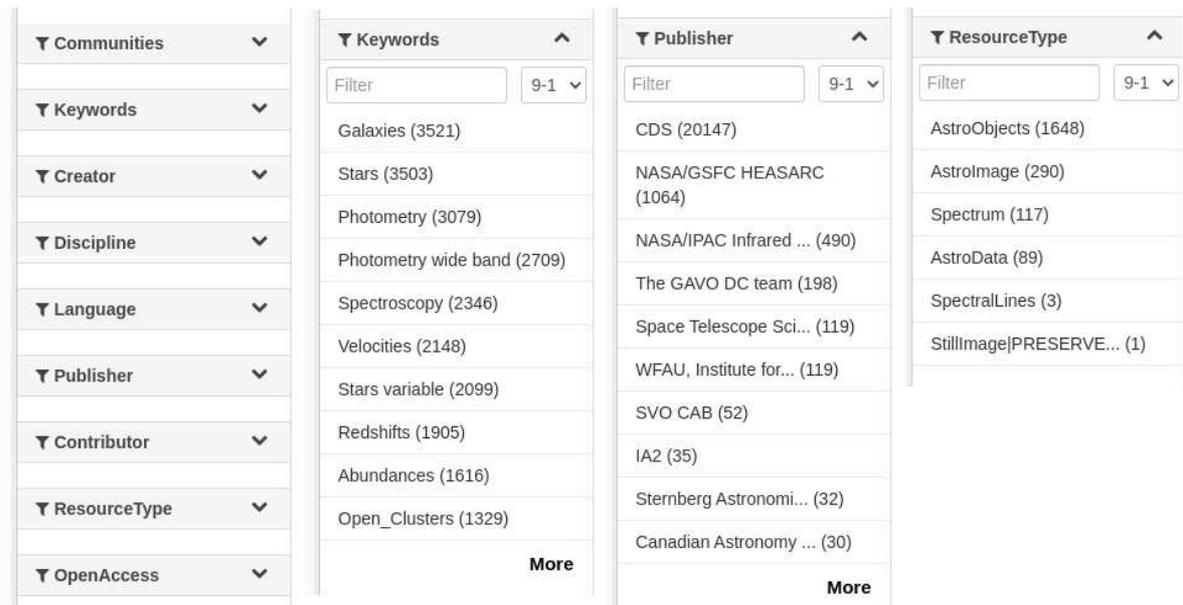}
   \end{tabular}
   \end{center}
   \caption[b2find-filters] 
   { \label{fig:b2find-filters} 
   The leftmost panel shows the features that B2FIND allows filtering upon. The other 
   three panels, Keywords, Publisher and ResourceType, show the available values for 
   each of the three features expanded (including the count of resources for each value).
}
\end{figure}

\begin{figure} [ht]
   \begin{center}
   \begin{tabular}{c} 
   \includegraphics[height=5cm]{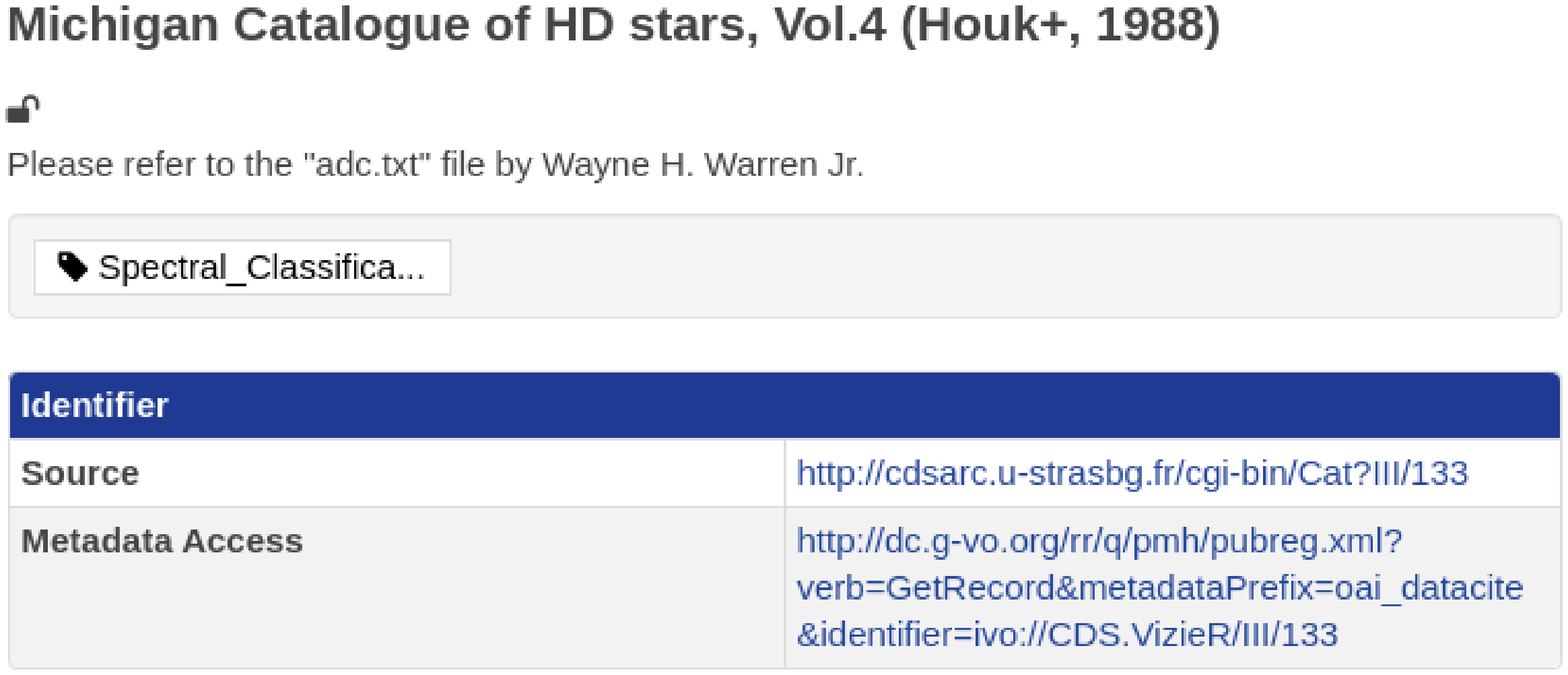}
   \end{tabular}
   \end{center}
   \caption[b2find-singleres] 
   { \label{fig:b2find-singleres} 
   B2FIND, single resource description. The Identifier metadata shows the Metadata Access that connects to the IVOA Registry content.
}
\end{figure}

The first activity towards integrating the IVOA architecture within EOSC has been to 
test the 
compatibility of and then include the IVOA Registry of Resources in the EOSC service 
catalogue. 
This activity started immediately at the beginning of the ESCAPE project, with the 
identification 
of contacts to check the technical details and the compatibility of metadata models. 
Contact was 
established with B2FIND, and the the resources of the IVOA Registry were then actually 
included 
into the B2FIND catalogue\footnote{This inclusion is visible at 
\url{http://b2find.eudat.eu/group?q=ivoa} (or \url{http://b2find.eudat.eu/group/ivoa)},
and illustrated by Figures~\ref{fig:b2find-search}, \ref{fig:b2find-results}, 
\ref{fig:b2find-filters} \& \ref{fig:b2find-singleres}.}, taking advantage
of the fact that the IVOA Registry of resources is 
OAI-PMH compliant. This activity is described in [\citenum{2019arXiv191108205M}].

Reports on the relevant 
activity during the IVOA Registry Working Group sessions in the IVOA meetings are 
captured in the 
Milestones documents M20 and M21 of ESCAPE CEVO ([\citenum{escape-m20}] \& 
\citenum{escape-m21}]). The IVOA Relational Registry Schema standard interface (RegTAP 
[\citenum{2019ivoa.spec.1011D}]) proved to be an efficient
tool for this task. The technical bases for B2FIND-IVOA interaction are the OAI-PMH 
transport 
protocol (already in use in both communities) and the DataCite metadata schema, which 
is an 
extension to the IVOA registry protocol. This extension is available on the GAVO (UHEI 
managed) 
full registry, which enabled the records to be harvested by B2FIND.

Figure~\ref{fig:b2find-search} shows how to start a quick-look of VO resources within 
EUDAT. A simple search driven by the
IVOA word yields the results page shown in Figure~\ref{fig:b2find-results}.

The plain result listing of the IVOA Registry content in B2FIND shown in 
Fig.~\ref{fig:b2find-results} can then be 
filtered by features, that are taken from the metadata elements of the resources 
description. As 
shown in Fig.~\ref{fig:b2find-filters}, the mapping of the metadata content between 
VOResource metadata and B2FIND metadata works well on the high-level description 
(\textit{Publisher} or \textit{Keywords} in Fig.~\ref{fig:b2find-filters}) 
but starts to be less informative the more one enters into the domain specific details (e.g., \textit{ResourceType}).

The filtering applied by B2FIND resembles the basic requirements answered by the IVOA 
Registry -- the fact that the metadata schema of the IVOA Registry includes the Dublin 
Core is an asset in that
respect -- but does not allow for the subsequent machine-actionable interaction that 
the standardised
interfaces provided by the IVOA standards allow. This is, of course, a known barrier 
when matching 
domain and general purpose repositories, because the depth of insight given by the 
descriptive 
metadata depends a lot on the semantic content of the metadata documents. Part of this 
is taken 
into account by the Metadata Access direct reference to the IVOA Registry full record (Fig.~\ref{fig:b2find-singleres}).

Further work towards cross-disciplinary integration has involved a study of cross 
domain user 
stories for resource discovery. The latter is a challenging task, which is not directly
included in
the ESCAPE work plan and its follow up will depend on potential connections to other 
cluster projects.

\subsubsection{Service on-boarding}
\label{subsubsec:onboarding}

Besides proving the technical feasibility of integrating the IVOA Registry content 
within the base 
repository structure of EOSC, our activities have also started to compare the concept 
of service 
onboarding, as described by EOSC, with the IVOA scenarios for service description and 
the 
deployment architecture. The \textit{Adding a service to 
EOSC}\footnote{\url{https://indico.in2p3.fr/event/20005/\#18-adding-a-service-to-eosc}} 
report, presented at the 
CEVO Technology Forum 1 by André Schaaff (CNRS-ObAS), was a first attempt in this 
direction. It was used
to report the experience in the EOSC-Enhance survey. While continuously checking on the
EOSC updates about service onboarding, for instance in the EOSC-Hub Week 2020 
report\footnote{\url{https://www.eosc-hub.eu/eosc-hub-week-2020/agenda/service-onboarding-catalogue-services}}, 
a further attempt will be made through cross-project connection with
H2020 NEANIAS\footnote{\url{https://www.neanias.eu/}}.

\subsubsection{Semantics}
\label{subsubsec:semantics}

Another set of resources to be tested for inclusion in the EOSC services are the 
IVOA-standardised 
semantic vocabularies\footnote{\url{https://ivoa.net/rdf/}}. The activity of checking 
how this 
integration could be done has already started, however progress on the EOSC side is 
currently not 
defined well enough to provide a clear connection at this stage of the project.

\subsection{VO Service Portfolio}
\label{subsec:portfolio}

This sub-section reports on the status of activities for actually providing resources 
to create a 
portfolio of services dedicated to astronomy within the EOSC.
These activities are at an early stage, given the complexity of the EOSC architecture. 
The EOSC-Hub
Week 2020 showed an inclusion plan for national, cluster and domain managed 
repositories. In CEVO 
we are following the evolution of that harvesting scenario while keeping awareness of 
other 
astronomy efforts, like the Europlanet VESPA containerised solution using the DaCHS 
resource and 
service software.
The idea of operating a custom marketplace for astronomy is probably better pursued in 
connection 
with ESCAPE WP3. This solution will be followed up.

\subsection{EOSC Hybrid Cloud Contribution}
\label{subsec:hybridcloud}

Another activity meant to integrate the VO architecture with the EOSC is through the 
federation of 
Research Infrastructure (RI) and connecting them to the EOSC itself. Here, the work has
progressed 
in connection with ESCAPE WP5 ESFRI Science Analysis Platform in two main directions.
On the one hand the UEDIN partner who leads the interactive data analysis work in WP5, 
is 
developing components to discover and access data analysis facilities based on agreed 
standardised 
APIs between the facility providers. To do so they are working on developing use cases,
notebooks 
and containers for the WP5 platform. Specific requirements for astrophysics sub-domains
(time domain, radio astronomy, \ldots) are collected and worked upon by all WP4 
partners.
On the other hand work is progressing for a possible implementation of the IVOA VOSpace
standard [\citenum{2018ivoa.spec.0621G}], the IVOA interface to distributed storage, to
contribute as a service to the ESCAPE/EOSC Cloud solution, possibly on top of the
ESCAPE Data Lake (WP2). A solution has been identified, both by 
INAF and UEDIN, to test VOSpace on top of RUCIO. It will be discussed with WP2.

\subsection{Service Containerisation}
\label{subsec:containers}

A final activity taken up to integrate and contribute to the EOSC cloud investigated 
the containerisation of web applications and other VO software to be integrated into 
the ESCAPE software repository. This activity will see a follow-up with WP3 Open-source
scientific Software and Service Repository, and the harmonisation of these containers 
with the science platform will see an interaction among WP4, WP5 and WP2 in the second 
half of the project.

\section{PROJECT INTERNAL \& EXTERNAL INTERACTIONS}

\subsection{Internal Cross-WP Interactions}
\label{subsec:crossWPintercation}

Progress on the integration activity described above has involved building connections 
between the WPs of the ESCAPE project.

\begin{figure} [ht]
   \begin{center}
   \begin{tabular}{c} 
   \includegraphics[height=7cm]{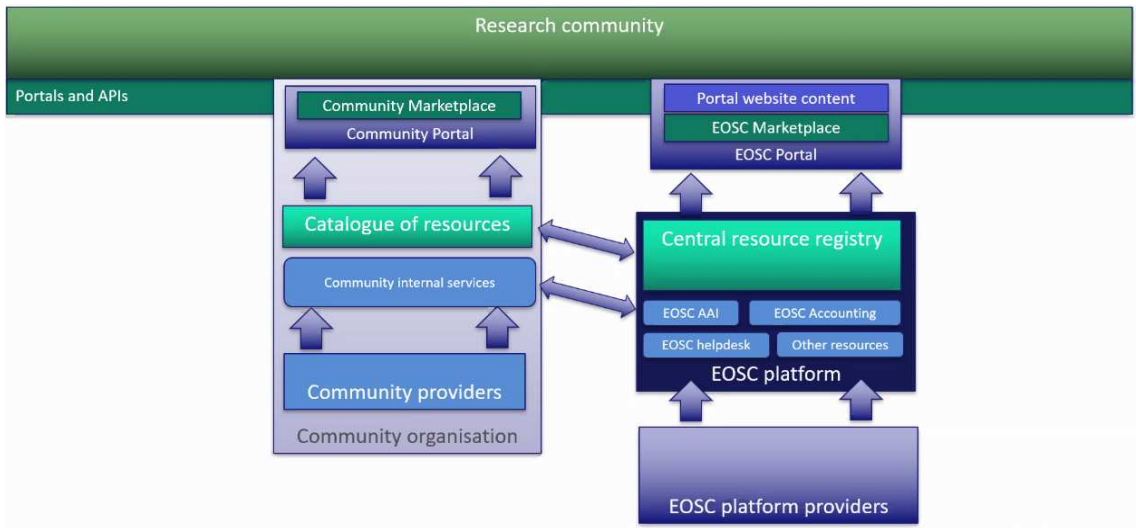}
   \end{tabular}
   \end{center}
   \caption[eosc-provide] 
   { \label{fig:eosc-provide} 
   EOSC architecture's foreseen catalogue integration. O. Appleton presentation
   at the EOSC-hub Week 2020.
}
\end{figure}

Activities led to provide software packages and services to the community through a 
FAIR repository
solution, which complement CEVO Task 4.1 to connect services directly to EOSC through 
the VO, are 
managed in ESCAPE WP3 Open-source scientific Software and Service Repository. Task 4.1 
got in 
connection with WP3 to contribute requirements to the ESCAPE repository. That 
repository, as shown 
during the EOSC-Hub Week 2020 (Fig.~\ref{fig:eosc-provide}), should flow into the 
general EOSC repository.
The WP3-WP4
collaboration should facilitate the contribution of VO resources to the ESCAPE 
repository and 
requirements gathering on media type annotations from the VO scenario.

Another point of connection is between CEVO EOSC integration activities and WP5 on 
containerisation
and hybrid cloud contributions. On the CEVO side this means porting IVOA derived 
technologies and 
having astronomy RIs implement or use them, while on the WP5 (ESAP) side it touches the
integration
of the Science Platform with the data resources. A specific aspect connected to this 
integration is
the discussion of the AAIs (Authentication and Authorization Infrastructures) and how 
they affect 
the existing landscape of VO resources and services.
Discussions held during the CEVO Technology Forum 1 and the ESCAPE Progress Meeting 
proved useful 
not only to make the above connections clear, but also to understand how the ESCAPE 
efforts in 
federating its RI resources could connect with the EOSC platform.

\subsection{Interactions with EOSC-related Projects}
\label{subsec:eoscinteraction}

In addition to connecting internally with other ESCAPE WPs and following the evolving 
FAIR best 
practices (e.g. through the RDA FAIR Data Maturity Model WG), our activities for 
integrating the VO
expertise within the EOSC landscape have also involved the interaction with other 
EOSC-related 
projects, particularly FAIRsFAIR and FREYA.

Mark Allen (CNRS-ObAS) is one of FAIRsFAIR Champions. He participated actively in the 
Second 
FAIRsFAIR Synchronisation Force Workshop, which was held as a series of sessions in 
April-June 
2020, during which he represented ESCAPE. The meeting was aimed at identifying the 
on-going 
activities related to the Action Plan of the “Turning FAIR into Reality” report, and to
identify 
gaps and possible additions to the Action Plan. Another FAIRsFAIR activity in which 
CEVO 
participated actively is the two workshops on metadata catlogue integration (9 
September and 11 
October 2020), and its follow-up DDI-CODATA Workshop (20 November 2020). This allowed 
us to explain
the IVOA metadata framework, and to participate in the discission on the possible 
generic solutions
and how disciplinary frameworks could interface with them.

CEVO participated in a meeting organised by FREYA on persistent identfiers in research 
disciplines 
(5 August 2020), which allowed us to present the way the IVOA defines and uses 
identifiers.

One also has to cite here the active participation of CEVO in the tests and discussions
of the RDA 
FAIR Data Maturity Model. The tests were performed jointly by Task 4.1 and Task 4.2: 
they enabled 
an assessment of how the IVOA framework was aligning with the proposed RDA Maturity 
Model and to 
identify discrepancies. The conclusion is that we can live with the maturity criteria, 
but that 
there are issues with the priorities: in particular, there is no “Essential” priority 
on 
Interoperability, whereas it is an essential element for astronomers, hence in the VO. 
The tests and
the conclusions from them, were reported during the Technology Forum and the IVOA 
meeting in May 
2020. This work brings useful information for the integration of the VO-enabled 
resources in EOSC, 
since the FAIR Metrics proposed by the EOSC FAIR WG are a subset of the RDA FAIR Data 
Maturity Model.

Attendance at the EOSC related meetings has been valuable both to showcase astronomy 
interoperability solutions, to present the developing astronomy requirements for EOSC, 
to understand the requirements and solutions from other research domains, and to get 
knowledge of the 
work performed by other projects and how astronomy could fit in.

\section{ANALYSIS OF THE INTEGRATION STATUS}
\label{sec:analysis}

Considering the progress shown in Sections~\ref{sec:integration} and 
\ref{subsec:crossWPintercation}, the integration of the IVOA-based open 
architecture of standards for astronomy into the connected ecosystem of data resources and services
in the EOSC is identified as being mostly feasible, but there are still challenges that need to be 
tackled to actually bring this integration to a mature level.

A definitely positive conclusion regards the high level description of data holdings and browser 
based services, leveraging the heritage and work of communities like the Open Archive Initiative, 
foundations like DataCite, DOI, ORCID and similar. The careful initial choice of the IVOA community
to base its registry of resources on the generic OAI-PMH protocol, and to include the Dublin Core 
as mandatory elements of the metadata schema, has made it relatively simple to map the basic 
annotations and concepts of the IVOA into metadata models and universal identifiers. This made 
feasible the integration of the IVOA Registry into the EOSC repository provided by EUDAT that also 
relies on OAI-PMH and DataCite.

The challenge is more on the machine actionability of the discovery and access services attached to
the data resources. The VO Registry uses standardised URIs to annotate the standard interfaces as 
well as to identify the relationships among resources. It has discipline-specific metadata for the 
coverage (e.g., in spatial and temporal coordinates) and scope of the data resources. EOSC, being 
the portal environment meant to be more general and abstract in terms of research domains, has a 
more abstract approach to data access and thus makes it difficult to reproduce the machine 
actionability level available in the IVOA. On the other side, IVOA, having developed its own URI 
schema for identifiers (IVOID) lacks somewhat on the take up of more general identifiers for its 
resources, i.e. the DOI nowaday more commonly used (although DOIs are commonly implemented by data 
providers, in particular for citation in publications). Another example of a difference emerging 
from the domain versus general repository behaviour are the tesselletion solutions (HiPS 
[\citenum{2017ivoa.spec.0519F}] and 
MOC [\citenum{2019ivoa.spec.1007F}] IVOA standards, based on HEALPix) provided by the VO for sky positional
and time coverage 
of the observations. Those are a valuable support in filtering the growing number of registered 
available resources and could be a technical feature to be reported alongside the other metadata 
also in a general portal solution like the EOSC one, being the natural counterpart of the 
geolocation of general web or earthbound resources.

A more challenging integration task has been comparing the user authentication and authorization 
(A\&A) solutions available within the astrophysical community and the interface architectures made 
available by the projects developing the EOSC.

On the one hand, the standardisation efforts within IVOA regarding A\&A have not seen a high level 
of implementation. This is due to the public approach to data and service delivery that represents 
the majority of VO registered holdings, as well as to the lack of a safely funded, globally 
operating identity provider. Moreover, IVOA standards, and VO contents consequently, deal more with
resource filtering, operations on catalogues (i.e. metadata annotated listings of astrophysical 
sources) and direct access to datasets, with a relatively minor support for providing computing 
operations. This means that, besides being only partially concerned by data access policies because
of the public data and resources approach, the VO community has, up to now, not been concerned with
accounting, because storage and computational resources made available to provide astrophysical 
data holdings for the community are usually enough to make them available and usable. 

On the other hand, FAIRness and Openness requirements in the VO architecture were, since the 
beginning, answered by interface protocols that were meant to be machine consumable. Thus the A\&A 
stress was more on the credential delegation technical aspects than the actual identification of 
the user and the permissions given to them.

For these reasons, the AAI (Authentication and Authorization Infrastructure) solutions being 
developed by the EOSC projects were initially not well aligned with the requirements of the VO data
providers: because on the one hand they pointed towards a licensing system where the only possible 
solution for most heritage data was a plain CC-0 (and that also felt difficult to attach in some 
cases), and on the other hand they lacked the machine actionability that full credential delegation
would allow.

Things have progressed in that respect since the start of the ESCAPE project, which can be seen by 
comparing the outcomes of EOSC Symposium (November 2019)1 with those of the EOSC-Hub Week 2019 
(April 2019)2 on the AAI aspects. Convergence and interoperability of the main A\&A solutions 
provided by the EOSC services (EGI Check-In, EUDAT B2ACCESS, GEANT eduTEAMS) has been provided to 
the benefit of the users. First discussions and awareness of machine-to-machine credential 
delegation were raised.

\begin{figure} [ht]
   \begin{center}
   \begin{tabular}{c} 
   \includegraphics[height=7cm]{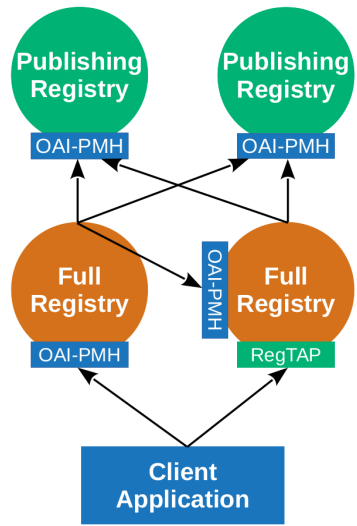}
   \end{tabular}
   \end{center}
   \caption[ivoa-registry] 
   { \label{fig:ivoa-registry} 
   IVOA Registry ecosystem: publishing, harvesting, cross-domain interoperability and domain specific interface solutions.
}
\end{figure}

There is also a difference in the way data resources and services are included in the VO Registry 
(and thus made available to VO-aware users and tools) with respect to what happens currently in the
EOSC onboarding scenario.
On the IVOA side a data/service provider has to be aware of the available typed-resources and how 
to describe them. The steps to perform then to publish a resource can be summarised as:

\begin{itemize}
    \item identify (or implement) a publishing registry and create an authority resource in it. This means 
providing an identifiable (it will have its own IVOID) root resource under which all the resources 
provided will be placed and could be uniquely identified by their ID;
    \item generate the appropriate (VO)Resource descriptions for its data collections and services and 
publish them through the registry.
\end{itemize}

The publishing registry, exposing its OAI-PMH interface, will be harvested by the full registries 
that form the actual resource repository known as the IVOA Registry. The data collections and 
services are made available to the VO community when the publishing registry is harvested. The 
above steps can be fully automated and human interaction is more on the operational side to check 
the health status of the Registry as a whole, following up on corrupted records, invalid resource 
descriptors, and unreachable resources.

Figure~\ref{fig:ivoa-registry} illustrates the way the IVOA Registry works. It shows how the 
OAI-PMH interface allows for
resource harvesting among the different registry instances and how the dedicated RegTAP 
solution 
can be used by applications (and proves more actionable). RegTAP [5] defines an interface for 
searching the resource metadata qtored in the registry using the IVOA Table Access Protocol. 
The 
same RegTAP interface is the one that facilitated the IVOA Registry mapping to the B2FIND 
repository.

On the EOSC side, the service onboarding solution is based on human interaction through
web forms and contact points until the resource/service is validated and released through the 
portal.

This difference might be a temporary one while EOSC tools and architecture become more mature.
It cannot be considered a challenge from the technical point of view because, as reported
above for 
the registry integration, the metadata models and FAIR approach are comparable. It is, 
however, a 
potential hurdle which could produce a slower take up of astrophysical resources within the 
EOSC 
ecosystem when progress will have been made on the service onboarding in EOSC.

Another challenging difference between the VO landscape and the EOSC proposed architecture is 
related to how service provision is handled. 

On the IVOA side, the contribution of data collections, services and other resources, is 
provided 
by the deployment of resources and services following the agreed standards (and often by 
autonomous
systems). This leads to an ecosystem where the validation of the deployed resources in the VO 
is an
operational activity focusing only on providing reliable resources to the end user.

In the EOSC architecture, where the governance is different, the data providers have to go 
through 
a different process, where metadata standards and interfaces are not automatically a means of 
creating an interoperable solution, and thus the onboarding process fills this gap by a 
preliminary 
validation of the resources that tends more to a preservation of the content and access points
to 
information holdings, rather than being a check of actionable interoperability.

The challenge here is to find a way to automatise as much as possible the on-boarding process 
for a
resources that are already registered in the IVOA, to reduce the amount of additional effort 
required on the providers side: the domain-specific technical one for the VO landscape, the 
general-purpose EOSC repository controlled validation on the other.

Last but not least important, a topic where integration is required is the interpretation on 
the 
FAIR principles applied to data resources and services already living in the VO landscape. 
This is 
an activity that crosses borders in CEVO between tasks 4.1 and 4.2 and a direct connection to
FAIR 
initiatives at the EOSC, but also global, level.
The IVOA vision, which has been developed starting in 2001-2002, was already following the
FAIR 
principles, since its aim was to enable astronomers to find, access and interoperate data, a
reuse 
capability being provided by the usage of a common format including data and metadata, and to 
provide seamless access to astronomical resources. of the concept of openness of data
resources and
services and always worked keeping in mind machine actionable interoperability. It can be seen
from the IVOA architecture schema shown in Fig.~\ref{fig:vo-fair}. IVOA developed a
disciplinary framework of 
FAIR practices which fulfils the community science needs but does not follow step by step all
the 
fine grained details of the FAIR principles, as for instance defined in the RDA FAIR Data
Maturity 
Model [\citenum{rda-fairdm-wg}].
The activities to be brought on, in connection to EOSC and the other EOSC-related project will
require testing the VO FAIR-ness with respect to EOSC FAIR Metrics and to continue on the work
already started for the RDA FAIR Data Maturity Model.
	
\section{NEXT STEPS}
\label{sec:next}

The integration of the VO architecture into the EOSC seems not an impossible task. Still it 
requires to tackle a few challenges that are currently discussed within the EOSC development 
bodies and projects.

From the point of view of the astrophysical community the main concern seems to be about how 
the existing machine-actionable and data providing semi-automated solution can be replicated in the EOSC without having to duplicate efforts in management and research technologies.

The effect of facing a multi-disciplinary environment has proven a challenge mostly because 
the astrophysical community concentrated on its science needs with respect to data sharing and
reuse, which led it to skip hurdles like the authentication and authorization of users to 
access discovery and retrieval service. This allowed for a direct path in maturing a re-usable
and interoperable paradigm of the VO that consequently brought into its architecture the 
findability, accessibility and interoperability aspects, thus completing the FAIR principles 
scenario in an operational framework at an early stage. 

Within ESCAPE, considering the progress made so far, the goal for the second part of the 
project  would be to investigate a solution for ESFRIs and RIs within the EOSC (and/or a 
domain specific, ESCAPE provided, cloud) to build their data resources and services adopting a
VO aware vision and, as seamlessly as possible, find those same resources available in EOSC 
itself and able to use the other resources provided within the EOSC such as computing 
resources.

To do so the challenges to tackle are currently the following:
\begin{itemize}
    \item attach data holdings to resource descriptions (not only PIDs for metadata);
    \item provide relationships among data resources and attached services;
    \item test, and make easier, the integration of VO-enabled data providers managed resources in the portal;
    \item deal with current AAI services and solution without preventing public content access;
    \item work at providing actionability for domain specific services and annotations through the EOSC portal;
    \item evaluate the position of VO-enabled data with respect to EOSC FAIR Metrics.
\end{itemize}

All the above challenges, building on the activities reported in this deliverable, will be 
tackled during the remaining part of the project. How far they proceed will depend on the 
actual status of the EOSC development and finalisation, and on the amount of resources 
necessary with respect to the tasks to perform.

\acknowledgments 

ESCAPE -- the European Science Cluster of Astronomy \& Particle Physics ESFRI Research 
Infrastructures -- has received funding from the European Union’s Horizon 2020 research and 
innovation programme under the Grant Agreement n. 824064.
MM acknowledges contacts and collaboration with the H2020 NEANIAS project in the person of Eva 
Sciacca.
MM, SB \& GT acknowledge the computing centre of the INAF -- Osservatorio Astronomico di 
Trieste ([\citenum{bertocco,taffoni}]).
\bibliography{11449-74} 
\bibliographystyle{spiebib} 

\end{document}